\documentclass[a4paper]{article}
\usepackage[dvips]{graphicx}
\usepackage{amsmath}
\usepackage{amssymb}
\usepackage{epsfig}
\usepackage{float}

\title{Space-time evolution induced by spinor fields with canonical and non-canonical 
kinetic terms}

\author{
Tomohiro Inagaki, Yury Rybalov\\
{\it Graduate School of Integrated Arts and Sciences, Hiroshima University,}\\
{\it Higashi-Hiroshima, 739-8521, Japan}\\[1mm]
Xinhe Meng\\
{\it Department of Physics, Nankai University, Tianjin, 300071, P.R.China}\\
}

\date{}

\begin{document}

\maketitle

\begin{abstract}
We study spinor field theories as an origin to induce space-time evolution.
Self-interacting spinor fields with canonical and non-canonical kinetic terms 
are considered in a Friedman-Robertson-Walker universe. The deceleration parameter 
is calculated by solving the equation of motion and the Friedman equation, 
simultaneously. It is shown that the spinor fields can accelerate and decelerate 
the universe expansion. To construct realistic models we discuss the contributions from the 
dynamical symmetry breaking.
\end{abstract}

\section{Introduction}
The acceleration periods of the universe evolution are one of 
the most important problems to be explored
in astroparticle physics today. Though the cosmological constant gives a 
simple solution to accelerate the universe expansion, a time dependent 
source is necessary to stop the inflation at the early universe and 
restart it at the present low energy scale.
There is no simple candidate to derive the 
accelerated expansion of the universe
in the Standard Model contents for particle physics. 
The problem calls for considerations to new field 
theory models which may describe the 
acceleration periods of the cosmic expansion.

Various models beyond the Standard Model have been studied to include 
candidates 
which accelerate the expansion of the universe.
The simplest model can be constructed by 
an inflaton like scalar field
which is easy to treat in a curved spacetime. A non-vanishing potential 
energy for the scalar field
can induce the exponential expansion of the universe. 
On the other hand, a spinor field also has played an important role as a
gravitational source. An ordinary 
free fermion behaves as a matter field and decelerates the universe expansion.
Much interest has been demonstrated in order to extend the model for 
the spinor field in recent years.

The study of spinor fields in curved spacetime has a long history.
The Dirac equation was investigated for massless spinor fields in 
curved space-time more than 50 years ago \cite{Brill:1957fx}. 
The Einstein-Dirac equations were solved for a massive spinor field 
in a special anisotropic spacetime, Bianchi type I universe with a 
cosmological constant \cite{Henneaux:1980ft}.
The possibility to induce
a primordial inflation \cite{Armendariz-Picon:2003} and
the current expansion in a self-interaction spinor field \cite{Ribas:2005}
has recently been pointed out.
Similar self-interacting fields have been discussed as a 
kind of Inflaton \cite{Watanabe:2009nc} and 
Quintom \cite{Cai:2008gk,Wang:2009ae}.
They have been studied in a Bianchi type I framework
\cite{Saha:1996,Saha:2000nk,Saha:2006b}
and other anisotropic space-time \cite{Saha:2004b,Saha:2010}. 
Non-standard spinors are considered as a source for the current expansion 
in Refs.~\cite{Gu:2006kw,Boehmer:2007ut,Boehmer:2008rz,%
Gredat:2008qf,Boehmer:2008ah,Shankaranarayanan:2009sz,Boehmer:2009aw,%
Wei:2010ad,Boehmer:2010ma}. There is a possibility to cause 
late-time cosmic acceleration
by a spinor field through a coupling with the 
Brans-Dicke scalar field \cite{Liu:2010,Samojeden:2010}, a form 
invariance transformation \cite{Chimento:2007fx} and a non-minimal 
curvature coupling \cite{Ribas:2007qm}.

The class of dark energy models has been proposed in a scalar field 
with non-canonical kinetic terms, 
such as models named
k-inflation \cite{ArmendarizPicon:1999rj,Garriga:1999vw} or 
k-essence \cite{Chiba:1999ka,ArmendarizPicon:2000dh,ArmendarizPicon:2000ah}.
A similar extension is possible for a spinor field. 
The spinor counterparts of k-essence for scalar field are called 
f-essence \cite{Myrzakulov:2010du,Tsyba:2011ss} and g-essence 
\cite{Yerzhanov:2010mt}.
We would like to continue the study of such models on 
self-interacting spinor fields with non-canonical kinetic terms as well.
It should be noted that models with the square of an usual Dirac 
Lagrangian may be special cases of f-essence scenario
\cite{Rakhi:2009a,Rakhi:2009b}.

The scalar invariant constructed from two spinor fields dynamically 
develops a non-vanishing value in QCD like 
theories \cite{Nambu:1961,Gross:1974jv}. Then the chiral
symmetry for the spinor fields is broken.  It has played an 
essential role for the evolution of the universe. Only a little work
has been done to evaluate the dynamical symmetry breaking in 
a non-static spacetime. In de-Sitter space it is possible
to calculate the expectation value for the scalar invariant
based on the maximal symmetry of the 
space \cite{Inagaki:1995jp,Hashida:1999wb}. 
However, we would like to find solution with early- and late-time 
acceleration periods. For this purpose it is not avoidable to 
develop different procedure.

In this paper a spinor field is assumed as a gravitational
source. We confine ourselves, for simplicity, in a 
Friedman-Robertson-Walker (FRW) metric and evaluate the 
corresponding space-time evolution. In Sec.~2 we review the 
space-time evolution induced by a spinor field with an ordinary 
canonical kinetic 
term. A self-interacting spinor field is considered in curved space-time.
Evaluating the equation of motion for the spinor field and Einstein's
field equations, we show typical behaviors of the Hubble and 
deceleration parameters. It is quite general to introduce a 
non-canonical kinetic term in an effective model. We assume 
a simple form for the non-canonical kinetic term and evaluate
the influences on the space-time evolution in Sec.~3. 
In Sec.~4 we set the critical scale where the dynamical
symmetry breaking takes place by hand and discuss the critical 
behavior caused by the dynamical symmetry breaking. The space-time 
evolution may be suddenly modified
by a first order phase transition. 
Finally, some concluding remarks are given in Sec.~5.

\section{Spinor fields with a canonical kinetic term}
Before we study special features of the non-canonical
kinetic term for a spinor field, it is more instructive 
to discuss the space-time evolution induced by a spinor 
field with an ordinary canonical kinetic term. The 
space-time evolution is evaluated by solving the equation
of motion for the spinor field and Einstein's field 
equations, simultaneously. The variations of the action
give rise to these equations.

Here we start from the action,
\begin{equation}
S=\int d^4 x \sqrt{-g}\left({\cal L}_g+{\cal L}_\Lambda+{\cal L}_D\right) ,
\label{action}
\end{equation}
where ${\cal L}_g$ is the Einstein-Hilbert Lagrangian and ${\cal L}_\Lambda$ 
describes a cosmological constant term
\begin{equation}
{\cal L}_g = \frac{1}{16\pi G}R ,
{\cal L}_\Lambda = -\frac{1}{8\pi G}\Lambda ,
\end{equation}
with $R$ the curvature scalar and $G$ the Newtonian constant.
The Dirac Lagrangian ${\cal L}_D$ is given by
\begin{equation}
{\cal L}_D = \frac{i}{2}\left[\bar{\psi}\Gamma^\mu D_\mu\psi
    -(D_\mu\bar{\psi})\Gamma^\mu\psi\right]
  -m\bar{\psi}\psi-V(\bar{\psi}\psi) ,
\label{lag:D}
\end{equation}
where the covariant derivatives for spinor fields, $\psi$ and $\bar{\psi}$, 
are defined by
\begin{equation}
 D_\mu\psi \equiv \partial_\mu\psi-\Omega_\mu\psi ,
\end{equation}
and
\begin{equation}
 D_\mu\bar{\psi} \equiv \partial_\mu\bar{\psi}-\bar{\psi}\Omega_\mu .
\end{equation}
The spinor connection, $\Omega_\mu$, is written by the vierbein, ${e^\mu}_a$,
\begin{equation}
 \Omega_\mu=-\frac{1}{4}g_{\rho\sigma}
 ({\Gamma^\rho}_{\mu\delta}-{e_{\rho}}_ b\partial_{\mu}{e_\delta}^b)
 \Gamma^\sigma\Gamma^\delta ,
\end{equation}
and the Dirac matrices, $\Gamma^{\mu}$, are generalized in a curved spacetime, 
\begin{equation}
\Gamma^{\mu}\equiv {e^{\mu}}_a \gamma^a.
\end{equation}

Here we suppose that the main contribution as a gravitational
source comes from a non-vanishing expectation value for a 
scalar invariant, $\bar{\psi}\psi$. The potential 
$V(\bar{\psi}\psi)$ can be written as a function of
a composite operator $\bar{\psi}\psi$.
We noticed that only scalar invariants can develop non-vanishing 
vacuum expectation values to keep the Lorentz structures in Minkowski 
spacetime. It is not valid in a curved space-time. Thus the above 
assumption should be modified in a strongly curved space-time.

The equations of motion for $\psi$ and $\bar{\psi}$ are given 
by the variations of the action (\ref{action}) with respect to
$\bar{\psi}$ and $\psi$, respectively,
\begin{equation}
 -i(D_\mu\bar{\psi})\Gamma^\mu - m \bar{\psi} -\frac{dV}{d\psi}=0 ,
\label{Dirac:1}
\end{equation}
and
\begin{equation}
 i\Gamma^\mu D_\mu\psi - m \psi -\frac{dV}{d\bar{\psi}}=0 .
\label{Dirac:2}
\end{equation}
The variation of the action (\ref{action}) with respect to
the vierbein gives the Einstein's field equations,
\begin{equation}
  R_{\mu\nu}-\frac{1}{2}Rg_{\mu\nu}+\Lambda g_{\mu\nu}=
  -8\pi G \langle T_{\mu\nu} \rangle,
\label{em:ein}
\end{equation}
where the energy-momentum tensor, $T_{\mu\nu}$, is given by
\begin{eqnarray}
  T_{\mu\nu}&=&\frac{i}{4}\left[
  \bar{\psi}\Gamma_\mu D_\nu \psi + \bar{\psi}\Gamma_\nu D_\mu \psi
  -(D_\nu \bar{\psi})\Gamma_\mu \psi - (D_\mu \bar{\psi})\Gamma_\nu \psi
  \right]
\nonumber \\
&&-g_{\mu\nu} {\cal L}_D.
\label{st:fermion}
\end{eqnarray}
The space-time evolution is found by solving Eqs.~(\ref{Dirac:1}), 
(\ref{Dirac:2}) and (\ref{em:ein}) with the help of the above energy-momentum form.

We consider the case that the spinor field has a homogeneous and isotropic 
distribution and look for the solution in a FRW universe with 
a flatly spatial part. The FRW universe is defined by the metric
\begin{equation}
  ds^2=dt^2-a^2(t)(dx^2+dy^2+dz^2) .
\label{metric:FRW}
\end{equation}
The space-time evolution is described by the time development
of the scale factor $a(t)$. Under the FRW metric, (\ref{metric:FRW}),
the generalized Dirac matrices are given by
\begin{equation}
  \Gamma^0 = \gamma^0, \ \ \Gamma^i = \frac{1}{a(t)}\gamma^i,
\end{equation}
and the spinor connection reads
\begin{equation}
  \Omega_0 = 0, \ \ \Omega_i = \frac{1}{2}\dot{a}(t)\gamma^i\gamma^0.
\end{equation}

Since the spinor field is assumed to be homogeneous and isotropic, 
it is written as only a function of time. In this
case Eqs.~(\ref{Dirac:1}) and (\ref{Dirac:2}) are simplified to be
\begin{equation}
 \partial_0 \bar{\psi} +\frac{3}{2}H \bar{\psi} - im \bar{\psi}\gamma^0
 - i\frac{dV}{d\psi}\gamma^0=0 ,
\label{Dirac:3}
\end{equation}
and
\begin{equation}
 \partial_0 \psi +\frac{3}{2}H \psi + i\gamma^0 m \psi 
 + i\gamma^0 \frac{dV}{d\bar{\psi}}=0 ,
\label{Dirac:4}
\end{equation}
where $H$ is the Hubble parameter, $H\equiv \dot{a}(t)/a(t)$.
The derivative of the potential, $V(\bar{\psi}\psi)$, can be 
written as
\begin{equation}
  \frac{dV}{d\psi}=V' \cdot \bar{\psi} ,\ \ \ \ 
  \frac{dV}{d\bar{\psi}}=V' \cdot \psi.
\label{dV}
\end{equation}
From Eqs.~(\ref{Dirac:3}), (\ref{Dirac:4}) and (\ref{dV}) 
the potential dependence can be eliminated. Therefore the 
composite operator, $\bar{\psi}\psi$, satisfies
a simple equation of motion,
\begin{equation}
  (\partial_0 +3H) \bar{\psi}\psi =0.
\label{em:barpsipsi}
\end{equation}
This equation has a trivial solution, $\bar{\psi}\psi =0$.
A non-trivial solution of Eq.~(\ref{em:barpsipsi}) is
\begin{equation}
  \bar{\psi}\psi = \frac{C}{a^3(t)} ,
\label{sol:barpsipsi}
\end{equation}
where $C$ is an arbitrary constant parameter fixed by an initial 
condition \cite{Armendariz-Picon:2003,Ribas:2005,Cai:2008gk}. 
The composite operator, $\bar{\psi}\psi$, develops as an 
ordinary matter field. We noticed that the solution (\ref{sol:barpsipsi}) 
is obtained at the classical limit. It may be modified by radiative 
corrections for the potential and the energy-momentum tensor.
It should be also noticed that the solution is found for a spatially 
constant, $\bar{\psi}\psi$. However, we consider that the solution 
has something essential for the development of the composite 
operator and discuss the evolution of the space-time under 
the solution (\ref{sol:barpsipsi}).

In the FRW universe the energy-momentum tensor, $T^{\mu}_{\nu}$,
has diagonal form. It can be parametrized as
\begin{equation}
  \langle T^{\mu}_{\nu} \rangle \equiv \mbox{diag}(\rho, -p, -p, -p) ,
\end{equation}
where $\rho$ corresponds to energy density 
and $p$ pressure.
Thus the Einstein's field equations (\ref{em:ein}) read
\begin{equation}
  3\left(\frac{\dot{a}(t)}{a(t)}\right)^2-\Lambda = 8\pi G \rho ,
\label{ein:00}
\end{equation}
and
\begin{equation}
  -2\left(\frac{\ddot{a}(t)}{a(t)}\right)
  -\left(\frac{\dot{a}(t)}{a(t)}\right)^2+\Lambda = 8\pi G p .
\label{ein:ii}
\end{equation}

From Eq.~(\ref{st:fermion}) with the equation of motion 
(\ref{Dirac:3}) and (\ref{Dirac:4}) the energy density, $\rho$, 
and the pressure, $p$, can be expressed as functions of the
scalar invariant, $\bar{\psi}\psi$,
\begin{equation}
  \rho=m \langle \bar{\psi}\psi \rangle 
  +\langle V(\bar{\psi}\psi)\rangle,
\label{rho:1}
\end{equation}
and
\begin{equation}
  p=-\langle V(\bar{\psi}\psi) \rangle 
  +\langle V'(\bar{\psi}\psi) \cdot \bar{\psi}\psi \rangle .
\label{p:1}
\end{equation}
An explicit form of the potential is necessary to find the typical 
behavior of the space-time. It is assumed that the potential can be 
expanded in terms of the scalar invariant,
\begin{equation}
  \langle V(\bar{\psi}\psi) \rangle
  =\sum_{n=1}^{\infty}\alpha_n \langle \bar{\psi}\psi \rangle^{2n},
\label{expand:V}
\end{equation}
where $\alpha_n$ corresponds to a coupling constant for multi-fermion 
interactions. 
To avoid an unexpected divergence we consider only a positive $n$ in 
Eq.~(\ref{expand:V}).
Since the scalar invariant, $\bar{\psi}\psi$, is a dimension 3 operator, 
the potential (\ref{expand:V}) is not renormalizable.
We regard the potential as a low energy effective one comming from 
a more fundamental theory at a high energy scale. It is expected that the higher 
order terms of $\bar{\psi}\psi$ are suppressed at low energy. Below,
we cut the summation at $n_{max}$. We noted that the model is
a kind of four-fermion interaction models for $n_{max}=1$ which
is often used as a low energy effective model of QCD 
\cite{Nambu:1961,Gross:1974jv}.

Hence, the energy density, $\rho$, and the pressure, $p$, read
\begin{equation}
  \rho=m\langle \bar{\psi}\psi\rangle  +
  \sum_{n=1}^{n_{max}}\alpha_n \langle \bar{\psi}\psi\rangle^{2n},
\label{rho:2}
\end{equation}
and
\begin{equation}
  p=\sum_{n=1}^{n_{max}}\alpha_n (2n-1) \langle \bar{\psi}\psi\rangle^{2n}.
\label{p:2}
\end{equation}
For a trivial solution of Eq.~(\ref{em:barpsipsi}), 
$\langle \bar{\psi}\psi\rangle =0$, the right hand 
sides of the Einstein's field equations, (\ref{ein:00}) 
and (\ref{ein:ii}) vanish, which simply describes a well known de Sitte universe. 
We can recognize that
the solution corresponds to lepton and heavy quark fields. 
In our universe finite vacuum expectation values are not 
observed for the lepton and the heavy quark bilinears. 
These spinor fields have no role for the evolution of 
the universe. However, it is known that the composite 
operator, $\bar{\psi}\psi$, develops non-vanishing value 
for up and down quarks in the current universe. Thus we 
evaluate the space-time evolution under the solution 
(\ref{sol:barpsipsi}).
Inserting the solution (\ref{sol:barpsipsi}) into Eqs.~(\ref{rho:2}) 
and (\ref{p:2}), we rewrite the Einstein's field equations, 
(\ref{ein:00}) and (\ref{ein:ii}), as
\begin{equation}
  3\left(\frac{\dot{a}(t)}{a(t)}\right)^2-\Lambda = 
  8\pi G \left[m\frac{C}{a^3(t)} 
  +\sum_{n=1}^{n_{max}}\alpha_n \frac{C^{2n}}{a^{6n}(t)}\right],
\label{ein:00:a}
\end{equation}
and
\begin{eqnarray}
  &&-2\left(\frac{\ddot{a}(t)}{a(t)}\right)
  -\left(\frac{\dot{a}(t)}{a(t)}\right)^2+\Lambda 
\nonumber \\
  &&= 8\pi G \left[\sum_{n=1}^{n_{max}}\alpha_n (2n-1) 
  \frac{C^{2n}}{a^{6n}(t)}\right] .
\label{ein:ii:a}
\end{eqnarray}
As is known, these equations are not independent. 
Equation (\ref{ein:ii:a}) can be derived from 
Eq.~(\ref{ein:00:a}) plus the Bianchi identity. 
These equations are called 
Friedman equations.

The Hubble parameter ,$H$, then is given by
\begin{equation}
  H^2 = \frac{\Lambda}{3}
  +\frac{8\pi G}{3} \left[m\frac{C}{a^3(t)} 
  +\sum_{n=1}^{n_{max}}\alpha_n \frac{C^{2n}}{a^{6n}(t)}\right].
\label{hubble}
\end{equation}
The cosmological constant and the mass terms mainly contribute 
to the right-hand side of Eq.~(\ref{hubble}) at the low energy 
limit, $a(t)\rightarrow\infty$.
As is well-known, the composite operator, $\bar{\psi}\psi$, 
develops a negative expectation value for up and down quarks 
in the current hadronic phase. It means that the parameter, 
$C$, has to be negative for the light quarks. Thus the mass 
term in Eq.~(\ref{hubble}) is negative in the standard 
model of particle physics. 
On the other hand, the left-hand 
side of Eq.~(\ref{hubble}) has possessed a non-negative value. The 
cosmological constant is necessary to take the low energy 
limit as, $a(t)\rightarrow\infty$. Upper limit for the scale 
factor exists without the cosmological constant. It should
be noted that the parameter $C$ is assumed to be positive
in Refs.~\cite{Ribas:2005,Cai:2008gk}.

The deceleration parameter is written as
\begin{equation}
  q\equiv -\frac{\ddot{a}(t)}{H^2 a(t)}=\frac{1}{2}(1+3\omega),
\label{deceleration}
\end{equation}
where $\omega$ shows the equation of state of the universe,
\begin{eqnarray}
  \omega
  &\equiv& \frac{-\Lambda+8\pi G p}
  {\Lambda +8\pi G \rho}
\nonumber \\
  &=& \frac{\displaystyle -\Lambda+8\pi G 
  \left[\sum_{n=1}^{n_{max}}\alpha_n (2n-1) 
  \frac{C^{2n}}{a^{6n}(t)}\right]}{\displaystyle \Lambda
  +8\pi G \left[m\frac{C}{a^3(t)} 
  +\sum_{n=1}^{n_{max}}\alpha_n \frac{C^{2n}}{a^{6n}(t)}\right]}.
\label{eos}
\end{eqnarray}
An acceleration period is realized for a negative $q$. 
At the low energy limit, $a(t)\rightarrow\infty$, it
diverges for $\Lambda=0$ and approaches a negative unity for a 
non-vanishing $\Lambda$.

\begin{figure}[htb]
\begin{minipage}{0.49\hsize}
\begin{center}
\resizebox{0.95\textwidth}{!}{
\includegraphics{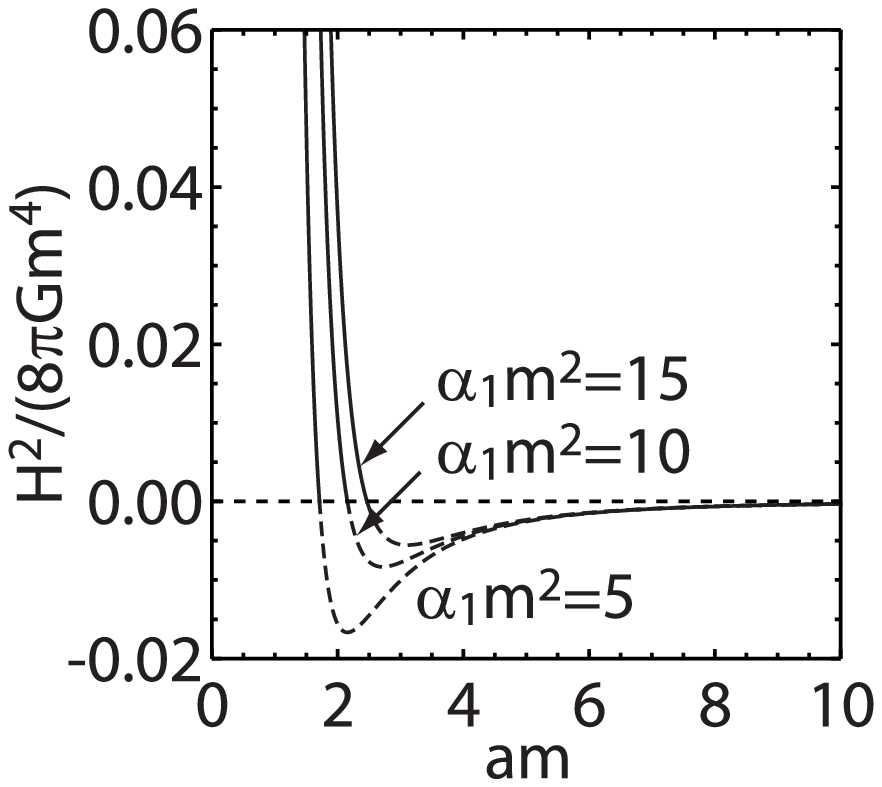}
}
\\
(a) $H^2/(8\pi G m^4)$
\end{center}
\end{minipage}
\begin{minipage}{0.49\hsize}
\begin{center}
\resizebox{0.86\textwidth}{!}{
\includegraphics{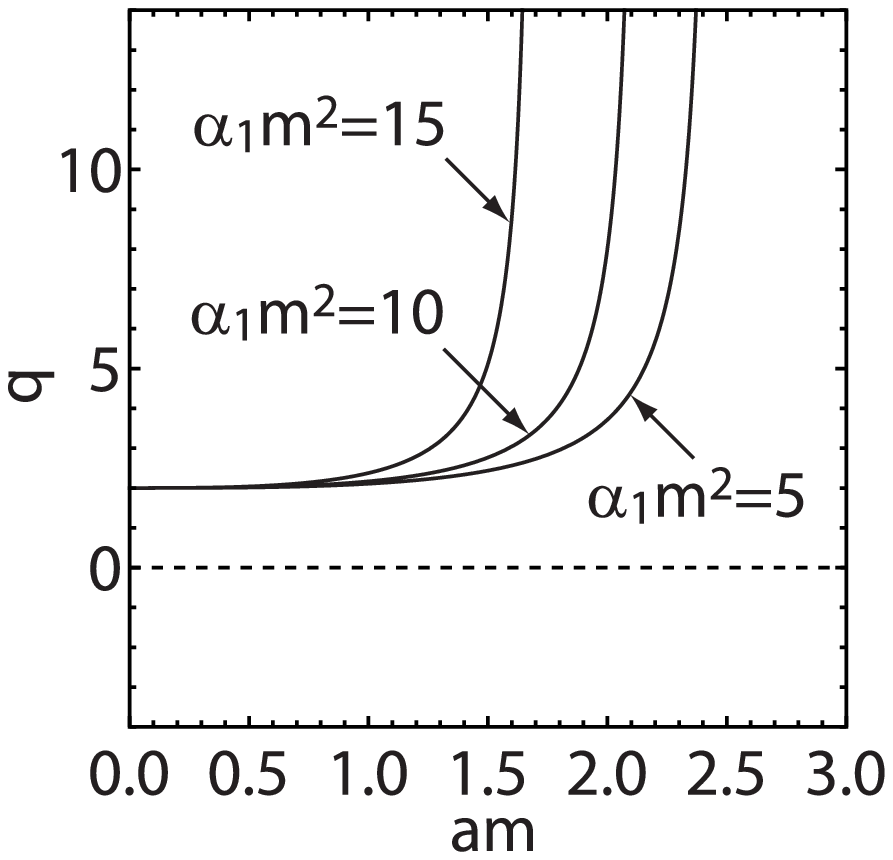}
}
\\
(b) $q$
\end{center}
\end{minipage}
\caption{Behavior of the Hubble and the deceleration 
parameters for $\Lambda=0$, $C=-1$, 
$\alpha_1 m^2=5, 10, 15$ and 
$n_{max}=1$.}
\label{Fig:hubble1}
\end{figure}

\begin{figure}[htb]
\begin{minipage}{0.49\hsize}
\begin{center}
\resizebox{0.95\textwidth}{!}{
\includegraphics{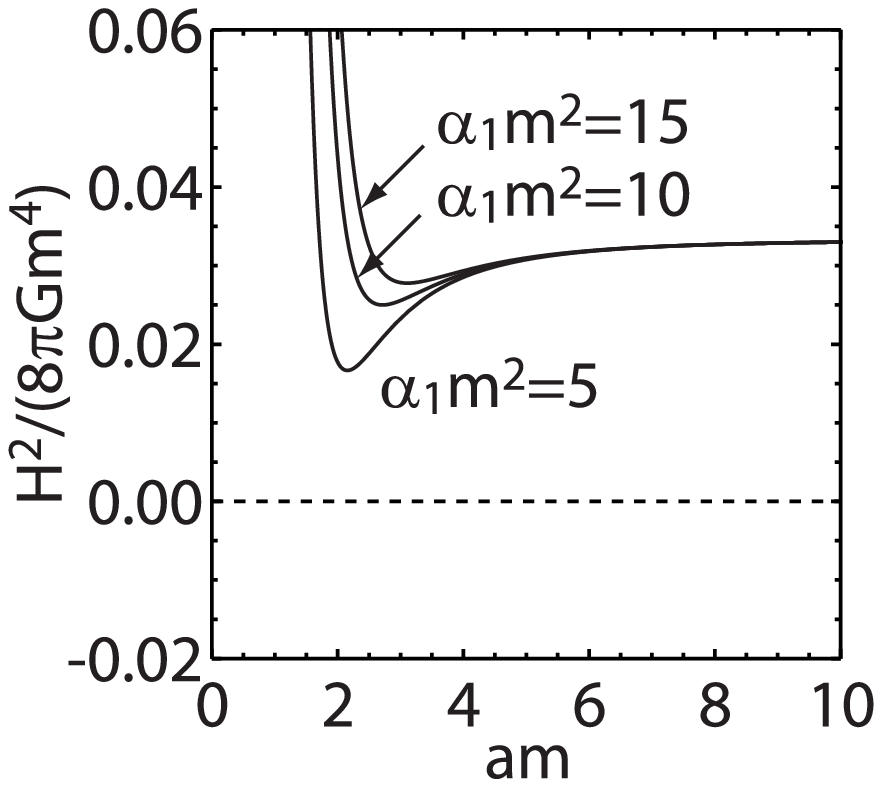}
}
\\
(a) $H^2/(8\pi G m^4)$
\end{center}
\end{minipage}
\begin{minipage}{0.49\hsize}
\begin{center}
\resizebox{0.86\textwidth}{!}{
\includegraphics{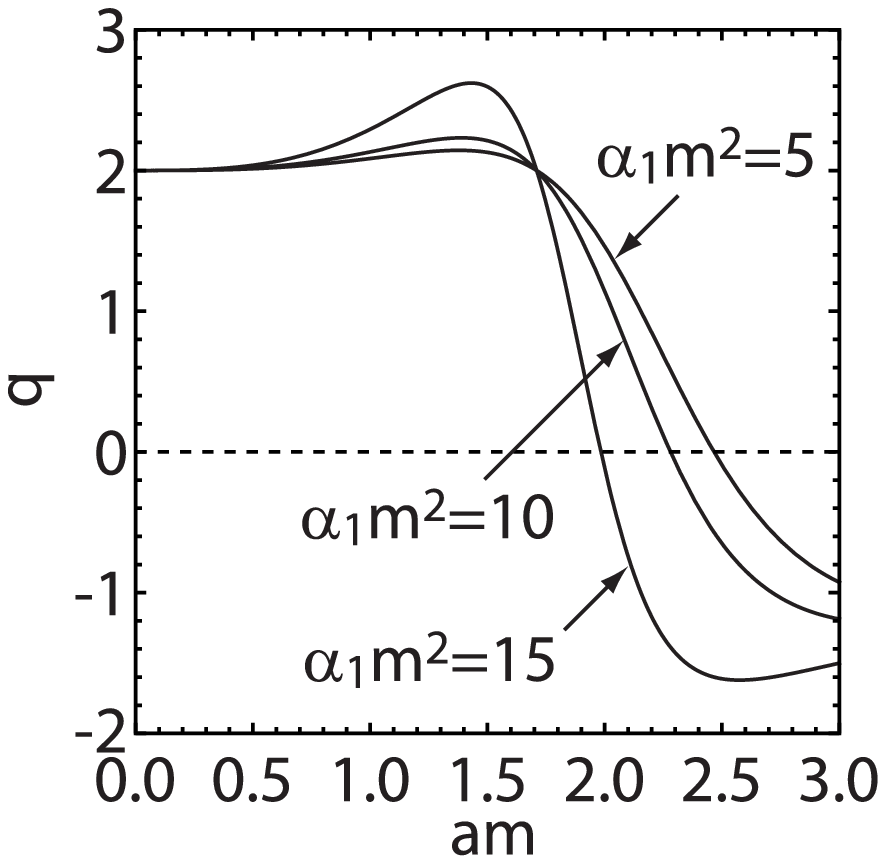}
}
\\
(b) $q$
\end{center}
\end{minipage}
\caption{Behavior of the Hubble and the deceleration 
parameters for $\Lambda/(8\pi G m^4)=0.1$, $C=-1$, 
$\alpha_1 m^2=5, 10, 15$ and $n_{max}=1$.}
\label{Fig:hubble2}
\end{figure}

Evaluating Eqs.~(\ref{hubble}) and (\ref{deceleration}) 
numerically, we show features of the spinor field as a
gravitational source. In numerical calculations all 
the mass scales are normalized by the fermion mass $m$. 
In Fig.~\ref{Fig:hubble1} typical behavior for the 
Hubble and the deceleration parameters are shown 
as a function of the scale factor $a$ for a four-fermion
interaction model without cosmological constant. An
eight-fermion interaction is introduced in
Fig.~\ref{Fig:hubble2}. The parameter, $C$, in 
Eq.~(\ref{hubble}) is negative for QCD-like theories.
The equation (\ref{hubble}) has a real solution for the 
Hubble parameter, only if the coupling constant for the 
higher dimensional operator, $\alpha_{n_{max}}$, takes a 
positive value. As is seen in Fig.~\ref{Fig:hubble1} 
(a), the right-hand side of Eq.~(\ref{hubble}) develops 
a negative value for a larger $a$. Since the Friedman 
equation has no real solution, the parameter range
for the dashed line is ruled out. There is an upper 
limit for the scale factor. In Fig.~\ref{Fig:hubble1} (b)
we observe that the deceleration parameter, $q$, 
blows up at the upper limit.

In Fig.~\ref{Fig:hubble2} we introduce a cosmological
constant term for a four-fermion interaction model. 
Since the cosmological constant term makes the 
right-hand side of Eq.~(\ref{hubble}) shift, it is
possible to cancel a negative contribution from the
fermion self interaction terms. In other words, a cosmological constant
like gravitational source is necessary to realize 
an open universe with a non-vanishing $\bar{\psi}\psi$.
In Fig.~\ref{Fig:hubble2} (b) the deceleration 
parameter converges to unity for a large $a$ limit.
Thus the late time universe can be accelerated
by the cosmological constant.

\begin{figure}[htb]
\begin{minipage}{0.49\hsize}
\begin{center}
\resizebox{0.86\textwidth}{!}{
\includegraphics{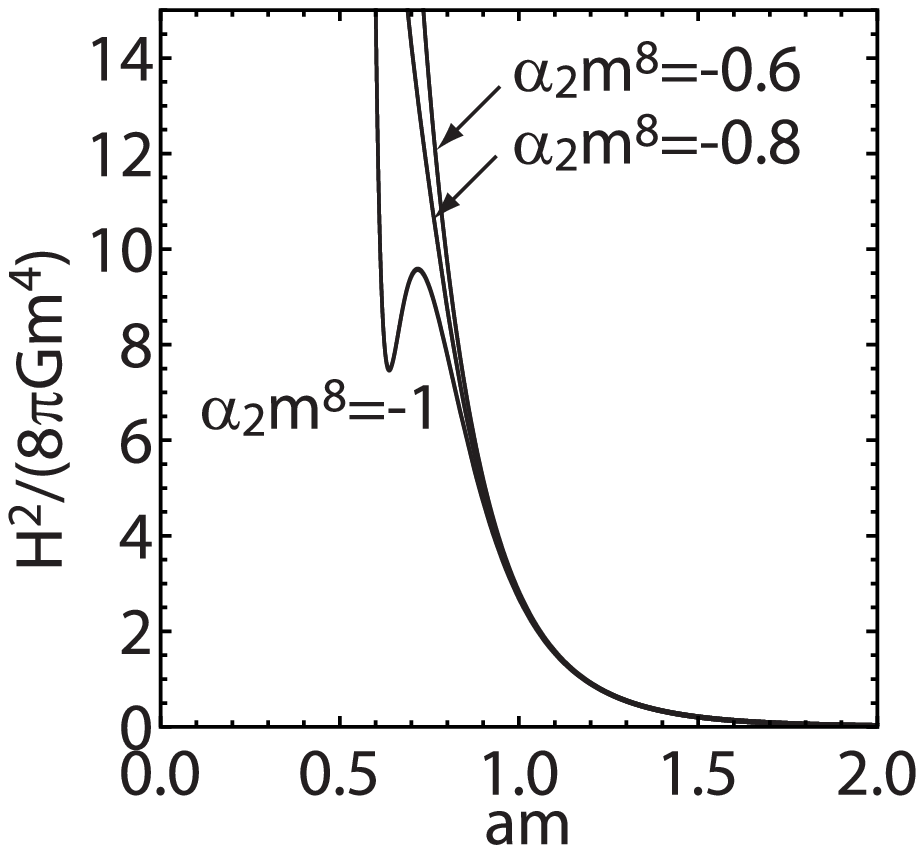}
}
\\
(a) $H^2/(8\pi G m^4)$
\end{center}
\end{minipage}
\begin{minipage}{0.49\hsize}
\begin{center}
\resizebox{0.95\textwidth}{!}{
\includegraphics{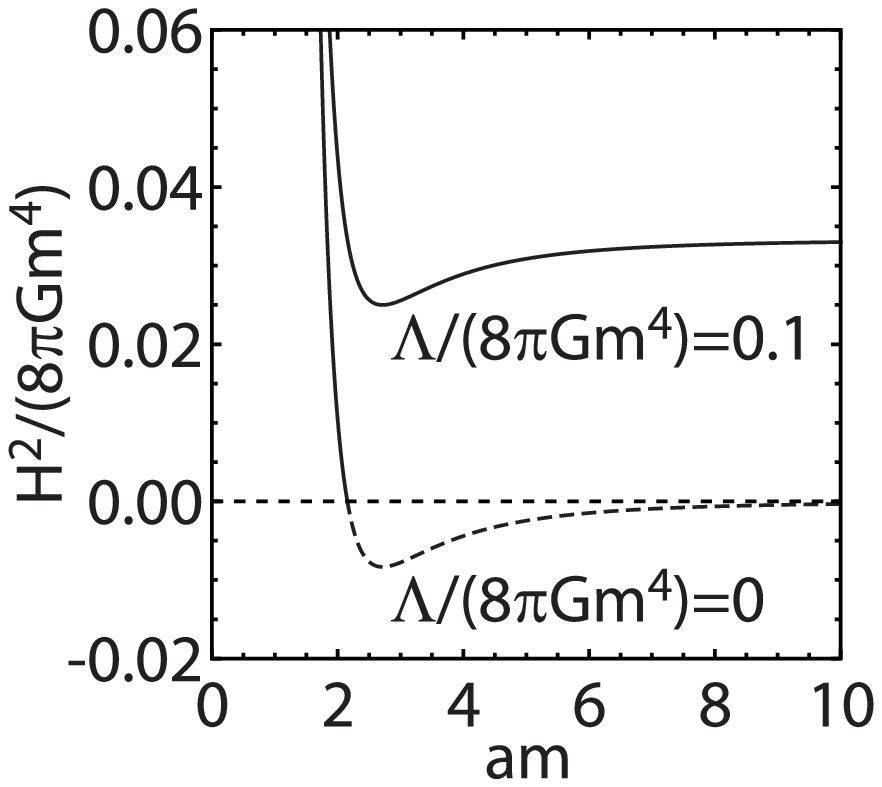}
}
\\
(b) $H^2/(8\pi G m^4)$
\end{center}
\end{minipage}
\caption{Behavior of the Hubble parameter for 
$\Lambda/(8\pi G m^4)=0, 0.1$, $C=-1$, $\alpha_1 m^2=10$, 
$\alpha_2 m^8=-0.6,-0.8,-1$, $\alpha_3 m^{14}=0.03$ and $n_{max}=3$.}
\label{Fig:hubble3}
\end{figure}

\begin{figure}[htb]
\begin{minipage}{0.49\hsize}
\begin{center}
\resizebox{0.86\textwidth}{!}{
\includegraphics{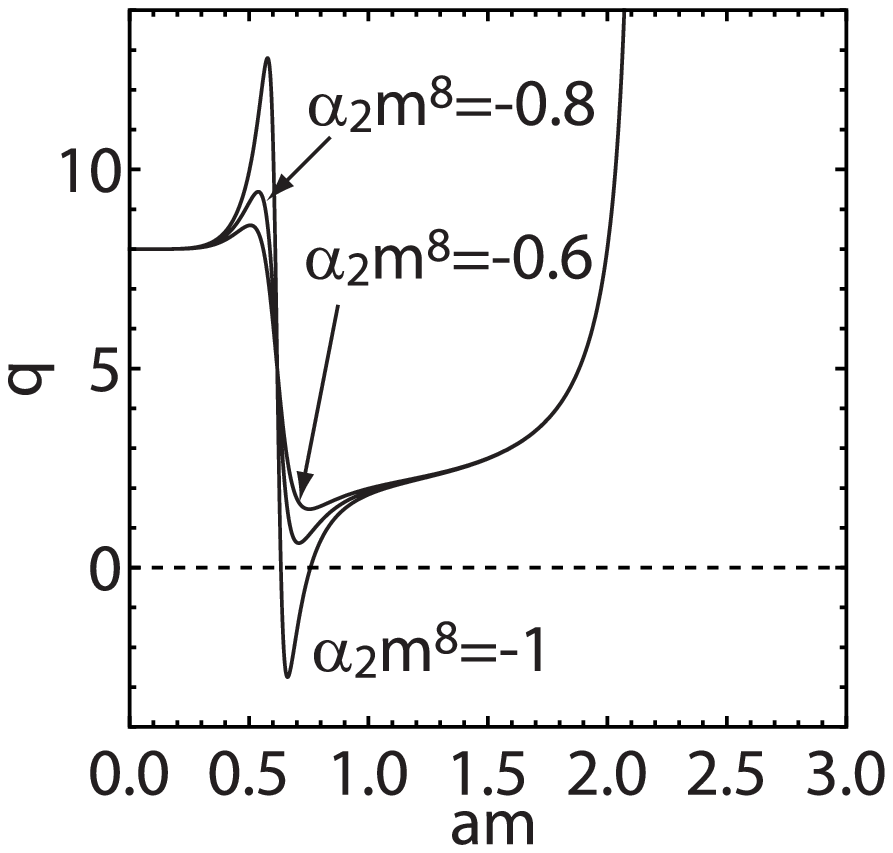}
}\\
(a) $\Lambda/(8\pi G m^4)=0$
\end{center}
\end{minipage}
\begin{minipage}{0.49\hsize}
\begin{center}
\resizebox{0.86\textwidth}{!}{
\includegraphics{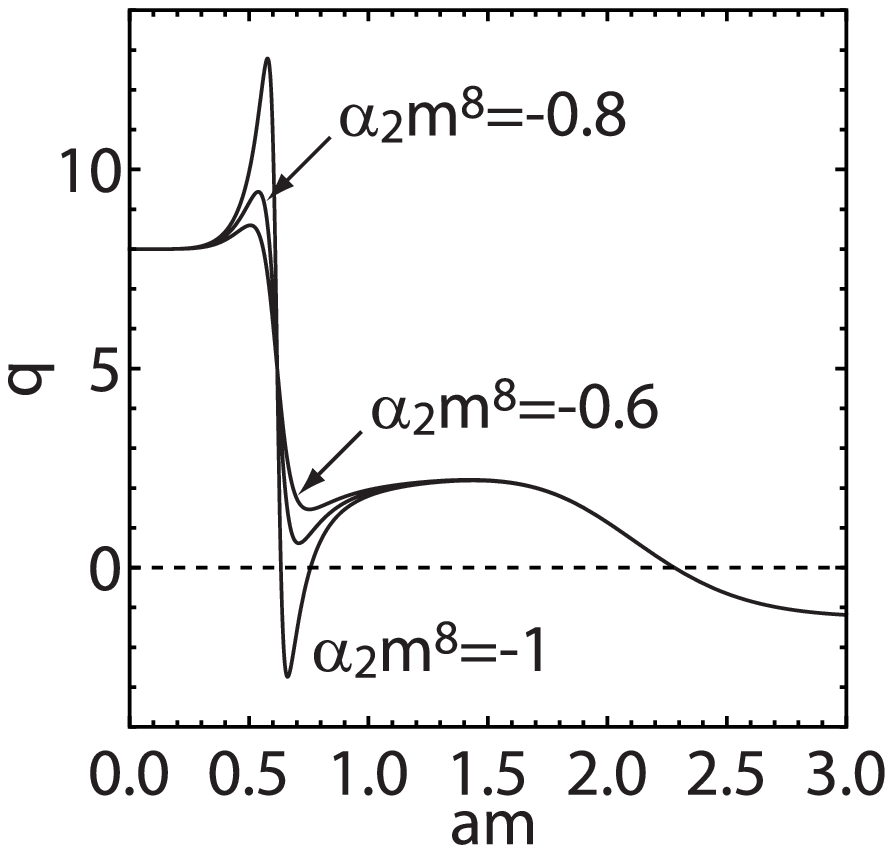}
}\\
(b) $\Lambda/(8\pi G m^4)=0.1$
\end{center}
\end{minipage}
\caption{Behavior of the deceleration parameter for 
$C=-1$, 
$\alpha_1 m^2=10$, 
$\alpha_2 m^8=-0.6,-0.8,-1$, $\alpha_3 m^{14}=0.03$ and $n_{max}=3$.}
\label{Fig:hubble4}
\end{figure}

For a small $a$ higher dimensional operators modify the 
behavior of the Hubble and the deceleration parameters.
Interesting behavior is found if one of $\alpha_n$
is negative. In Fig.~\ref{Fig:hubble3} we draw the behavior
of the Hubble parameter for a negative eight-fermion
coupling, $\alpha_2 <0$. We also introduce a positive
twelve-fermion coupling, $\alpha_3 >0$ to stabilize the
potential. A negative $\alpha_2$ modifies the behavior of 
Hubble parameter in Fig.~\ref{Fig:hubble3} (a). As is shown
in Fig.~\ref{Fig:hubble3} (b) the square of the Hubble 
parameter becomes negative without cosmological constant.
Thus the Friedman equation has no real solution on the
parameter range for the dashed line. 
This again means that 
the universe has an upper limit. The deceleration 
parameter blows up at the scale in Fig.~\ref{Fig:hubble4} 
(a). In the case $a_2m^8=-1$ we find an acceleration 
period, $q<0$, without any cosmological constant. 
In Fig.~\ref{Fig:hubble4} (b) it is found that the late 
time universe can also be accelerated by introducing 
cosmological constant.

\section{Non-canonical kinetic term}
In the previous section we only consider scalar type 
multi-fermion interactions. It is assumed that
such interactions are introduced in a field theory as 
low energy effective model stemming from a more fundamental 
theory at high energy scales. Varieties of interactions which keep 
the actual symmetry for the particle physics can be 
generated. A kinetic term for fermion may 
also be modified 
at low energy scale. Here we investigate a class of fermion 
field theories which are described by the Dirac Lagrangian 
with non-canonical kinetic terms,
\begin{eqnarray}
{\cal L}_D &=& \frac{i}{2}f(\bar{\psi}\psi)\left[\bar{\psi}\Gamma^\mu D_\mu\psi
    -(D_\mu\bar{\psi})\Gamma^\mu\psi\right]
\nonumber \\
  &&+m\bar{\psi}\psi-V(\bar{\psi}\psi),
\label{mod:D}
\end{eqnarray}
where the function $f(\bar{\psi}\psi)$ modifies the kinetic term.
If we set $f(\bar{\psi}\psi)=1$, the Lagrangian (\ref{lag:D}) is reproduced.
Here we consider a more general case, where the function $f(\bar{\psi}\psi)$
is given by a function of only a scalar invariant $\bar{\psi}\psi$.

The non-canonical kinetic terms modify the equations of motion 
for fermion fields, $\psi$ and $\bar{\psi}$, to
\begin{eqnarray}
&& \frac{i}{2} f'\bar{\psi}\left[\bar{\psi}\Gamma^\mu D_\mu\psi
    -(D_\mu\bar{\psi})\Gamma^\mu\psi\right]
-\frac{i}{2}(\partial_{\mu} f)\bar{\psi}\Gamma^{\mu}
\nonumber \\
&&-i f (D_\mu\bar{\psi})\Gamma^\mu 
+ m \bar{\psi} -V'\bar{\psi}=0 ,
\label{Dirac:5}
\end{eqnarray}
and
\begin{eqnarray}
&& \frac{i}{2} f'\psi\left[\bar{\psi}\Gamma^\mu D_\mu\psi
    -(D_\mu\bar{\psi})\Gamma^\mu\psi\right]
+\frac{i}{2}(\partial_{\mu} f)\Gamma^{\mu}\psi
\nonumber \\
&& +i f \Gamma^\mu D_\mu\psi
 + m \psi -V'\psi=0 ,
\label{Dirac:6}
\end{eqnarray}
where we write
\begin{equation}
f'\equiv \frac{d f(\bar{\psi}\psi)}{d (\bar{\psi}\psi)}.
\end{equation}
Derivatives of the potential, $V(\bar{\psi}\psi)$, and the 
function $f(\bar{\psi}\psi)$ can be written as
\begin{equation}
  \frac{dV}{d\psi}=V' \cdot \bar{\psi} ,\ \ 
  \frac{dV}{d\bar{\psi}}=V' \cdot \psi ,\ \ 
  \frac{df}{d\psi}=f' \cdot \bar{\psi} ,\ \ 
  \frac{df}{d\bar{\psi}}=f' \cdot \psi.
\label{der}
\end{equation}
From Eqs.~(\ref{Dirac:5}) and (\ref{Dirac:6}) with Eqs.~(\ref{der}) 
we obtain a simple equation of motion for the composite operator, 
$\bar{\psi}\psi$,
\begin{equation}
  \left[\left(f+{\bar{\psi}\psi}f'\right)\partial_0 +3fH\right] 
\bar{\psi}\psi =0.
\label{em:barpsipsi2}
\end{equation}
There is a trivial solution, $\bar{\psi}\psi =0$. To find a non-trivial solution 
we expand the function $f(\bar{\psi}\psi)$ in terms of the scalar invariant,
\begin{equation}
f(\bar{\psi}\psi)=\sum_{n} \beta_n(\bar{\psi}\psi)^{2n}.
\end{equation}
Thus the non-trivial solution of Eq.~(\ref{em:barpsipsi2}) is found to be
\begin{equation}
  f(\bar{\psi}\psi) \bar{\psi}\psi  = 
  \sum_{n} \beta_n(\bar{\psi}\psi)^{2n+1} = \frac{C}{a^3(t)} ,
\label{sol:barpsipsi2}
\end{equation}
where $C$ is a constant parameter.

\begin{figure}[htb]
\begin{minipage}{0.49\hsize}
\begin{center}
\resizebox{0.86\textwidth}{!}{
\includegraphics{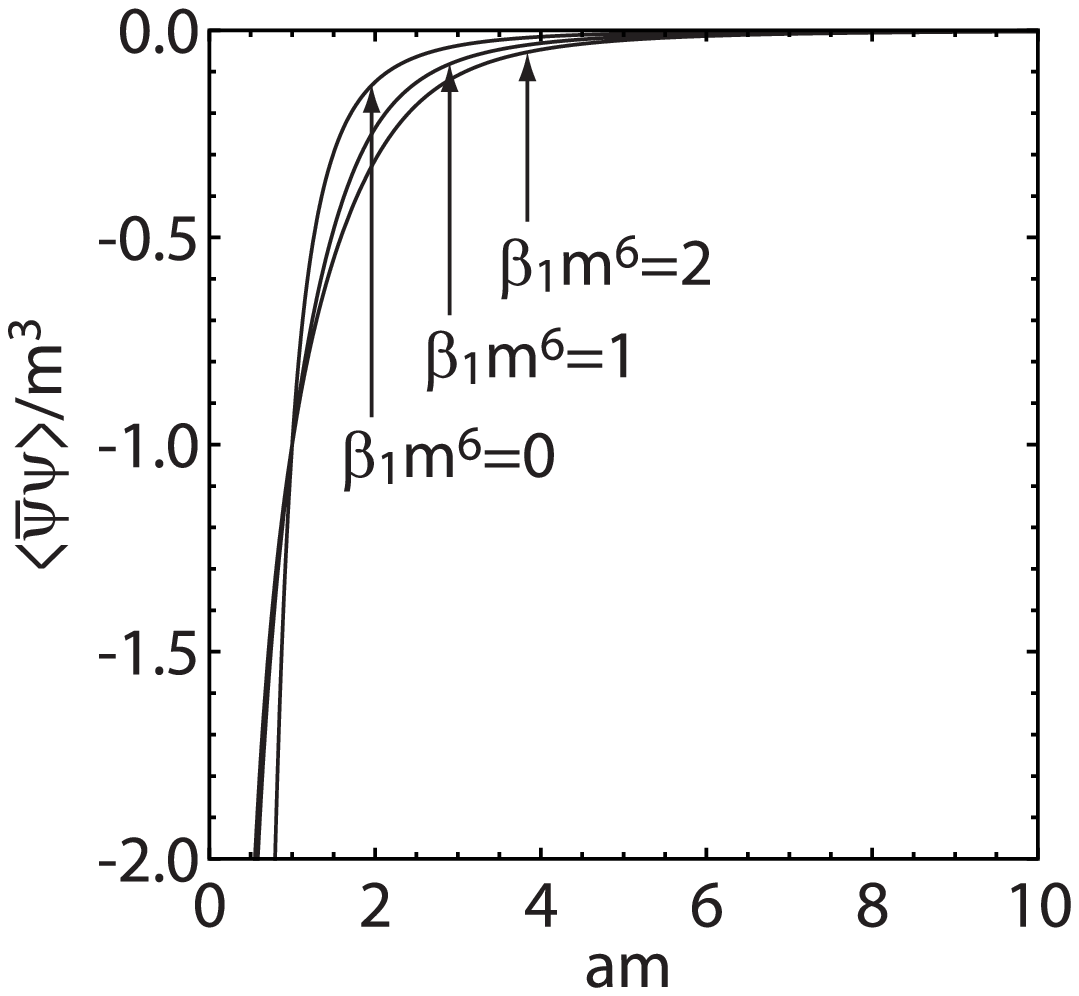}
}
\\
(a) $\beta_{-1}=\beta_{2}=0$
\end{center}
\end{minipage}
\begin{minipage}{0.49\hsize}
\begin{center}
\resizebox{0.86\textwidth}{!}{
\includegraphics{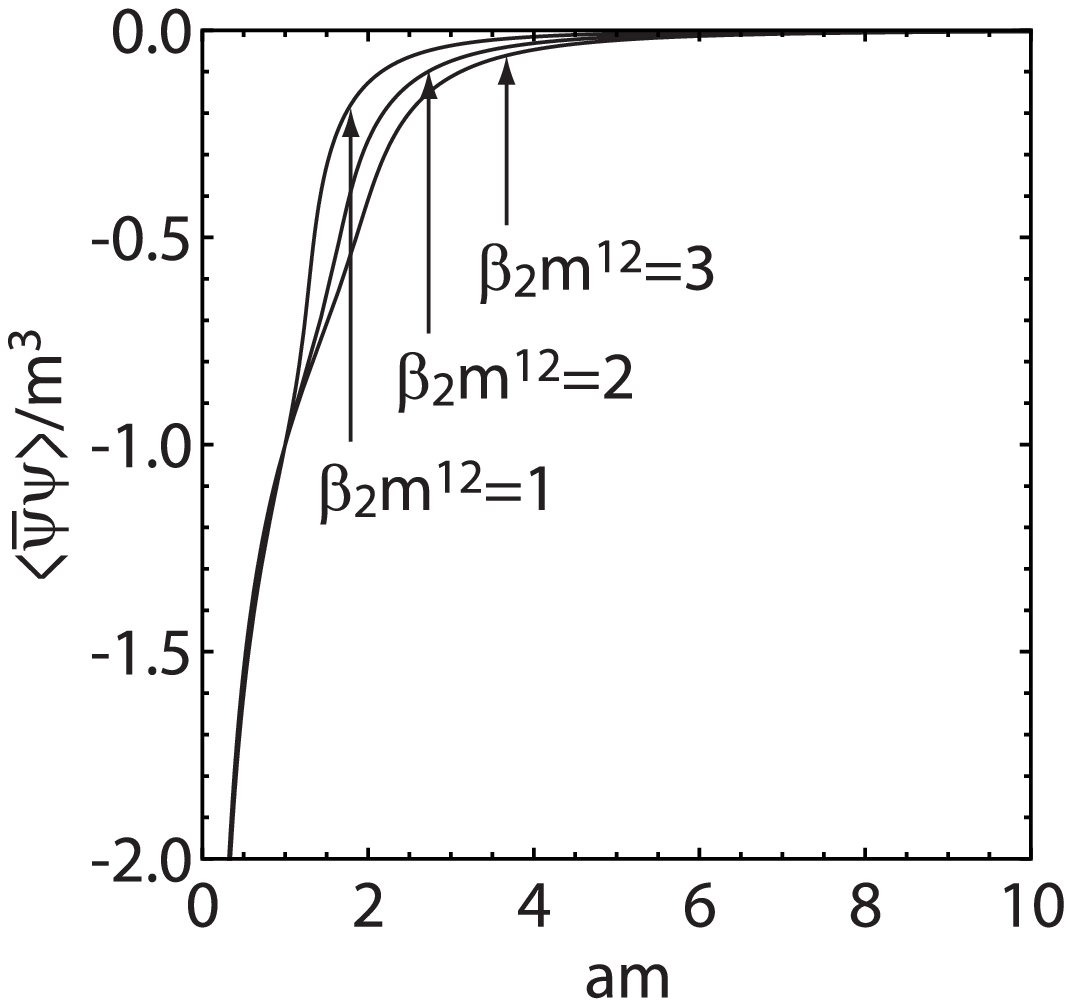}
}\\
(b)  $\beta_{-1}=0$, $\beta_1m^6=-1$
\end{center}
\end{minipage}
\vglue 5mm
\begin{minipage}{0.49\hsize}
\begin{center}
\resizebox{0.86\textwidth}{!}{
\includegraphics{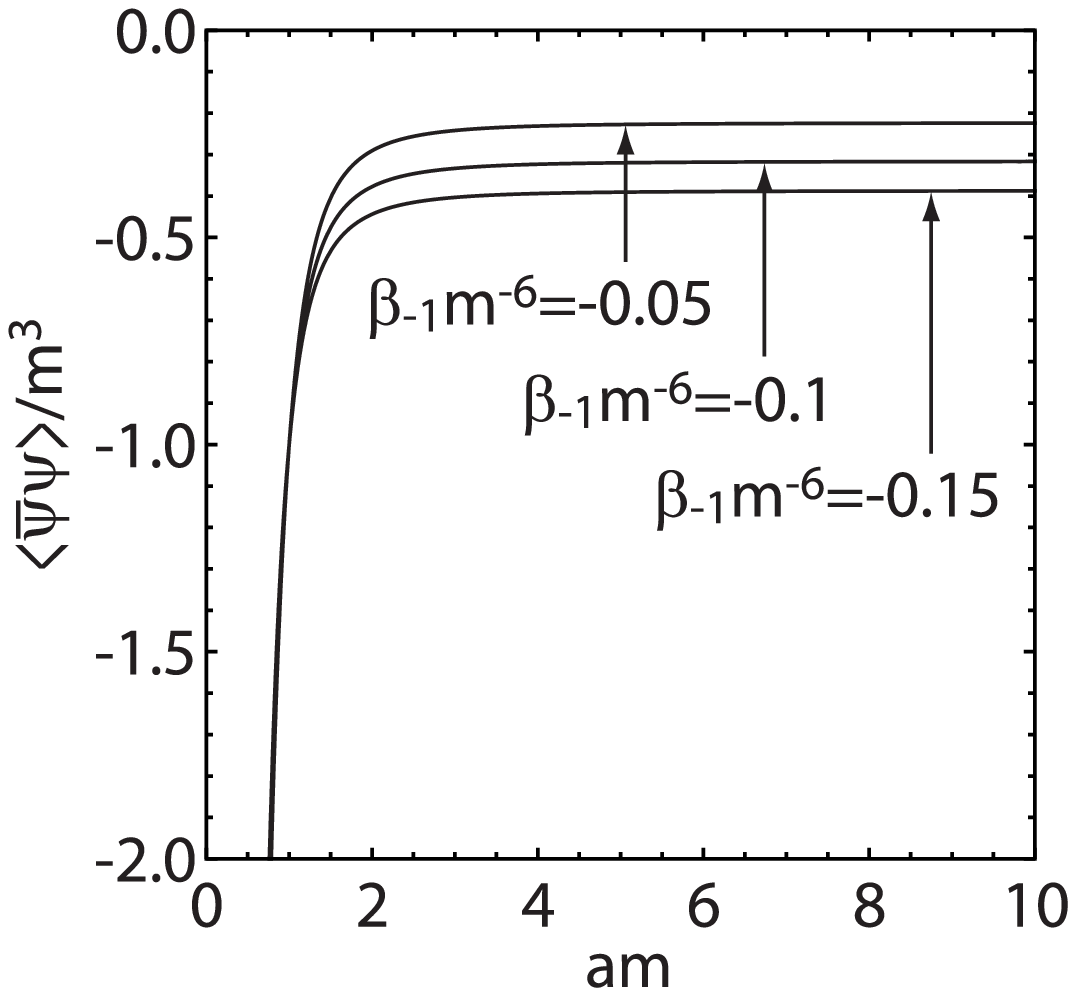}
}\\
(c) $\beta_1=\beta_{2}=0$
\end{center}
\end{minipage}
\begin{minipage}{0.49\hsize}
We set $\beta_0=1$ and $\beta_{n}=0$ for $n\neq -1, 0, 1, 2$.
\end{minipage}
\caption{Behavior of the expectation value
$\langle\bar{\psi}\psi\rangle$ for 
$\langle\bar{\psi}\psi\rangle (t=t_0)=m^3, a(t=t_0)=m^{-1}$.}
\label{Fig:fc}
\end{figure}

We numerically solve Eq.~(\ref{sol:barpsipsi2}) for a fixed $a(t=t_0)$,
$\langle\bar{\psi}\psi\rangle (t=t_0)$ and $\beta_n$. We draw 
typical behavior of $\langle\bar{\psi}\psi\rangle$ in Fig.~\ref{Fig:fc}
with $\langle\bar{\psi}\psi\rangle (t=t_0)=m^3$ and $a(t=t_0)=m^{-1}$.
The term, $\beta_1 (\bar{\psi}\psi)^2$, changes the slope for a small $a$.
If the coefficient, $\beta_1$, is negative, a positive $\beta_2$ is necessary
to satisfy Eq.~(\ref{sol:barpsipsi2}) for a large 
$|\langle\bar{\psi}\psi\rangle|$. 
In this case a more gradual slope
is observed for a larger $\beta_2$.
The term, 
$\beta_{n} (\bar{\psi}\psi)^{2n}$, with a negative $n$ 
diverges at the limit, $\bar{\psi}\psi\rightarrow 0$. 
Non-canonical kinetic term remains at a late time universe. 
Thus the term, $\beta_{-1} (\bar{\psi}\psi)^{-2}$, is less interesting.
In Fig.~\ref{Fig:fc} (c) a negative $\beta_{-1}$ raises up a value,
$|\bar{\psi}\psi|$, for a large $a$.

In the FRW universe the Einstein's field equations are given by Eqs.~(\ref{ein:00})
and (\ref{ein:ii}). 
For the Lagrangian (\ref{mod:D}) the energy-momentum tensor, $T_{\mu\nu}$, is 
modified to be
\begin{eqnarray}
  T_{\mu\nu}&=&\frac{i}{4} f(\bar{\psi}\psi)
 \left[
  \bar{\psi}\Gamma_\mu D_\nu \psi + \bar{\psi}\Gamma_\nu D_\mu \psi
  -(D_\nu \bar{\psi})\Gamma_\mu \psi \right.
\nonumber \\
  && \left. - (D_\mu \bar{\psi})\Gamma_\nu \psi
  \right]-g_{\mu\nu} {\cal L}_D. 
\nonumber \\
\label{st:fermion2}
\end{eqnarray}
From Eq.~(\ref{st:fermion2}) with the equation of motions 
(\ref{Dirac:5}) and (\ref{Dirac:6}) we find the same expression for the 
energy density, $\rho$, and the pressure, $p$, with Eqs.~(\ref{rho:2}) 
and (\ref{p:2}), respectively.
Thus the Friedman equations read
\begin{equation}
  3\left(\frac{\dot{a}(t)}{a(t)}\right)^2-\Lambda = 
  8\pi G \left(m\langle \bar{\psi}\psi\rangle  +
  \sum_{n=1}^{n_{max}}\alpha_n \langle \bar{\psi}\psi\rangle^{2n}\right),
\label{ein:00:2}
\end{equation}
and
\begin{eqnarray}
  &&-2\left(\frac{\ddot{a}(t)}{a(t)}\right)
  -\left(\frac{\dot{a}(t)}{a(t)}\right)^2+\Lambda 
\nonumber \\
  &&= 8\pi G \left(\sum_{n=1}^{n_{max}}\alpha_n (2n-1) \langle \bar{\psi}\psi\rangle^{2n}\right).
\label{ein:ii:2}
\end{eqnarray}
As is mentioned in the previous section, the Eq.~(\ref{ein:ii:2}) can
be derived from the Eq.~(\ref{ein:00:2}) with the help of continuous equation 
for the homogenous pressure and density from the cosmic fermionic components. 

\begin{figure}[htb]
\begin{minipage}{0.49\hsize}
\begin{center}
\resizebox{0.95\textwidth}{!}{
\includegraphics{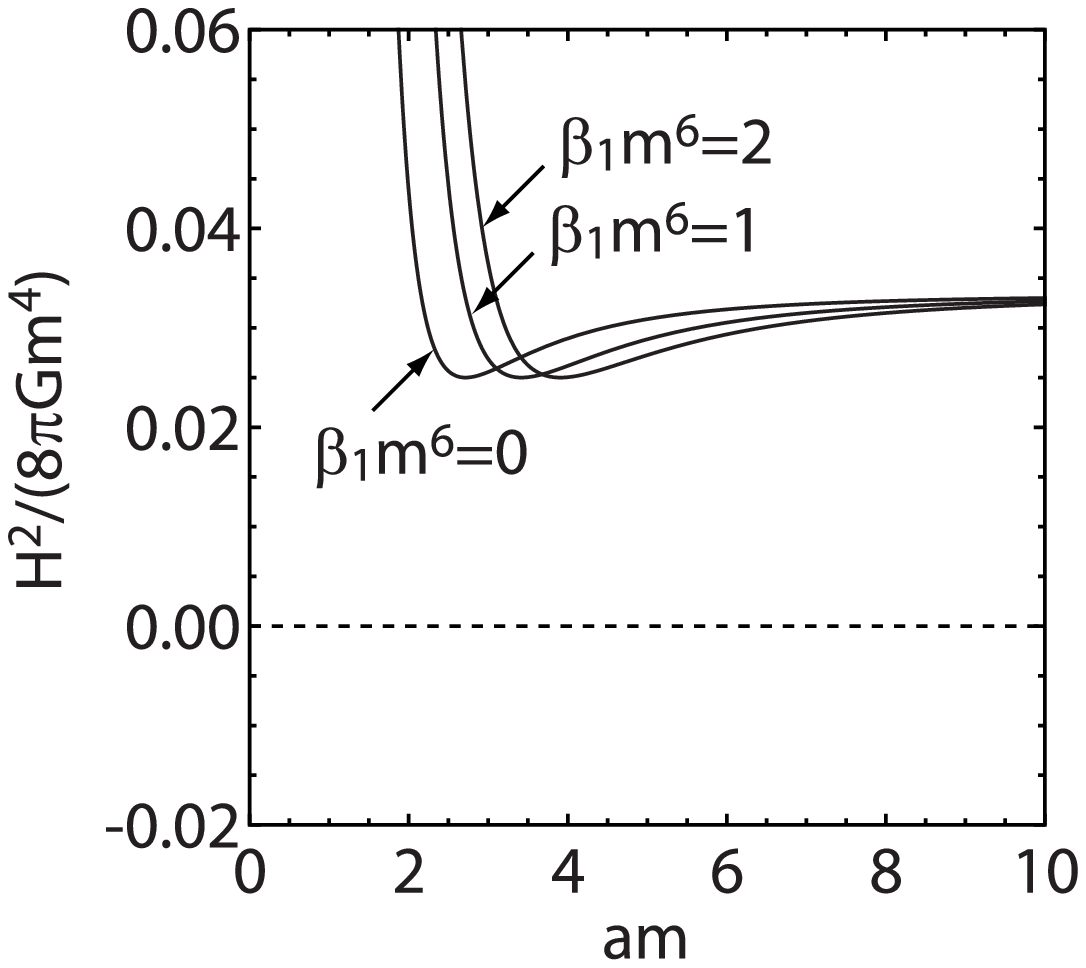}
}\\
(a) $H^2/(8\pi G m^4)$
\end{center}
\end{minipage}
\begin{minipage}{0.49\hsize}
\begin{center}
\resizebox{0.86\textwidth}{!}{
\includegraphics{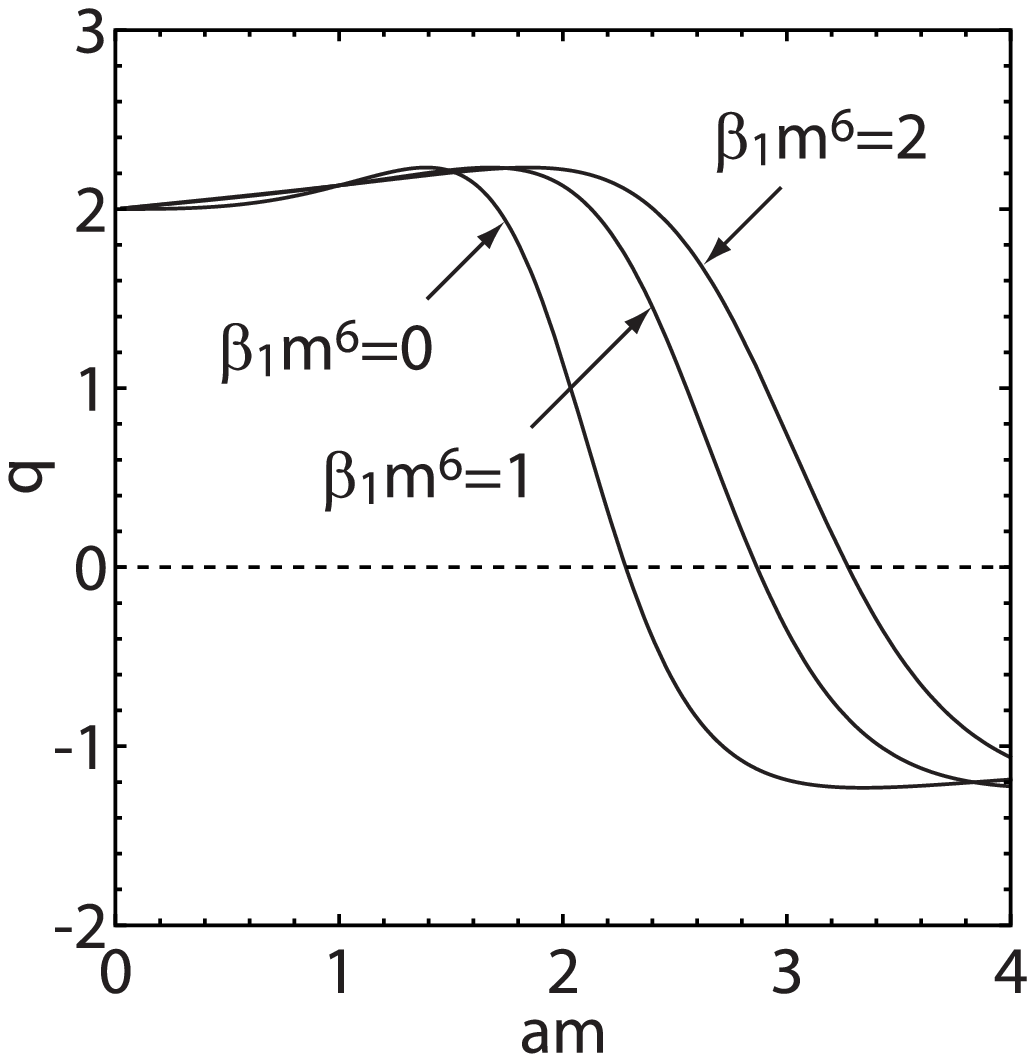}
}\\
(b) $q$
\end{center}
\end{minipage}
\caption{The behavior of the Hubble and the 
deceleration parameters for $\Lambda/(8\pi G m^4)=0.1$,
$\langle\bar{\psi}\psi\rangle (t=0)=m^3, a(t=0)=m^{-1}$, 
$\alpha_1 m^2=10$, $n_{max}=1$, $\beta_0=1$, 
$\beta_1m^6=0, 1, 2$ and $\beta_m=0$ for $m\neq 0,1$.
}
\label{Fig:hubble01}
\end{figure}

\begin{figure}[hbt]
\begin{minipage}{0.49\hsize}
\begin{center}
\resizebox{0.95\textwidth}{!}{
\includegraphics{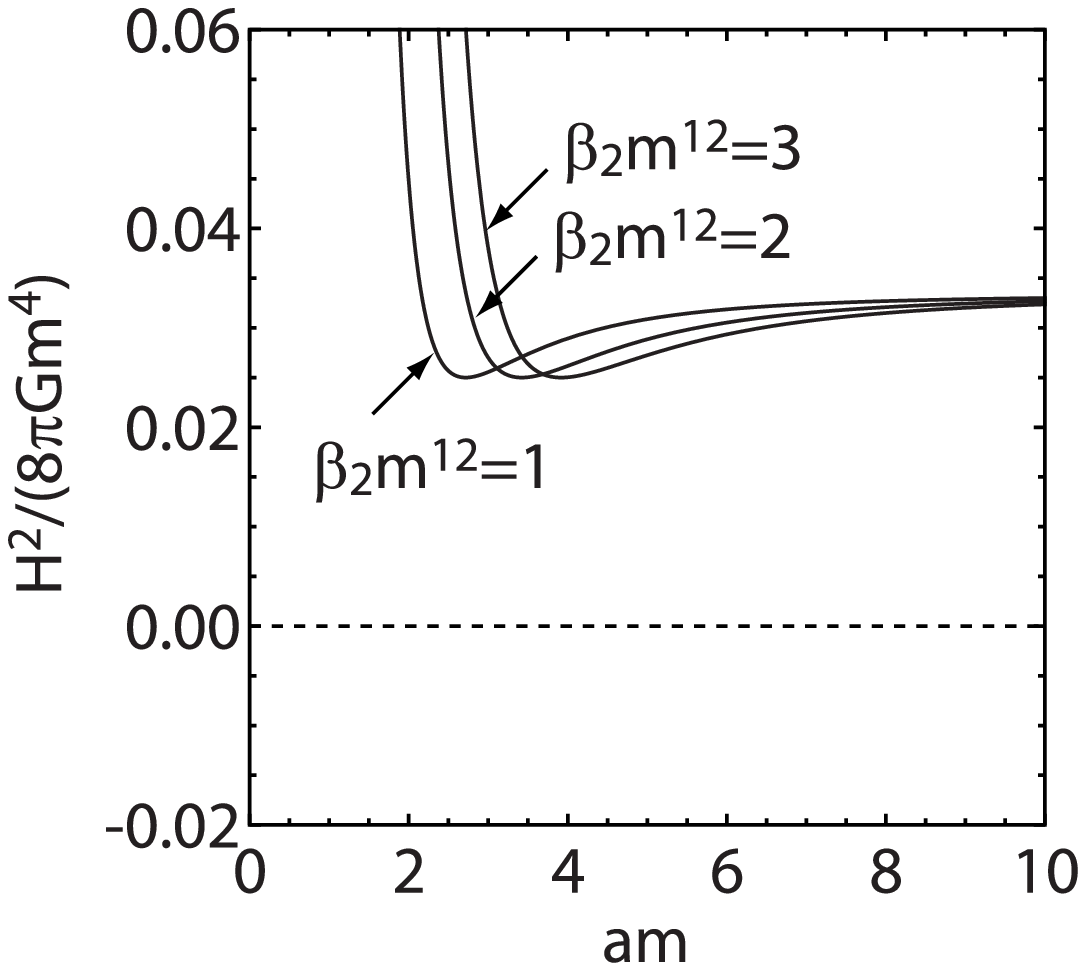}
}\\
(a) $H^2/(8\pi G m^4)$
\end{center}
\end{minipage}
\begin{minipage}{0.49\hsize}
\begin{center}
\resizebox{0.86\textwidth}{!}{
\includegraphics{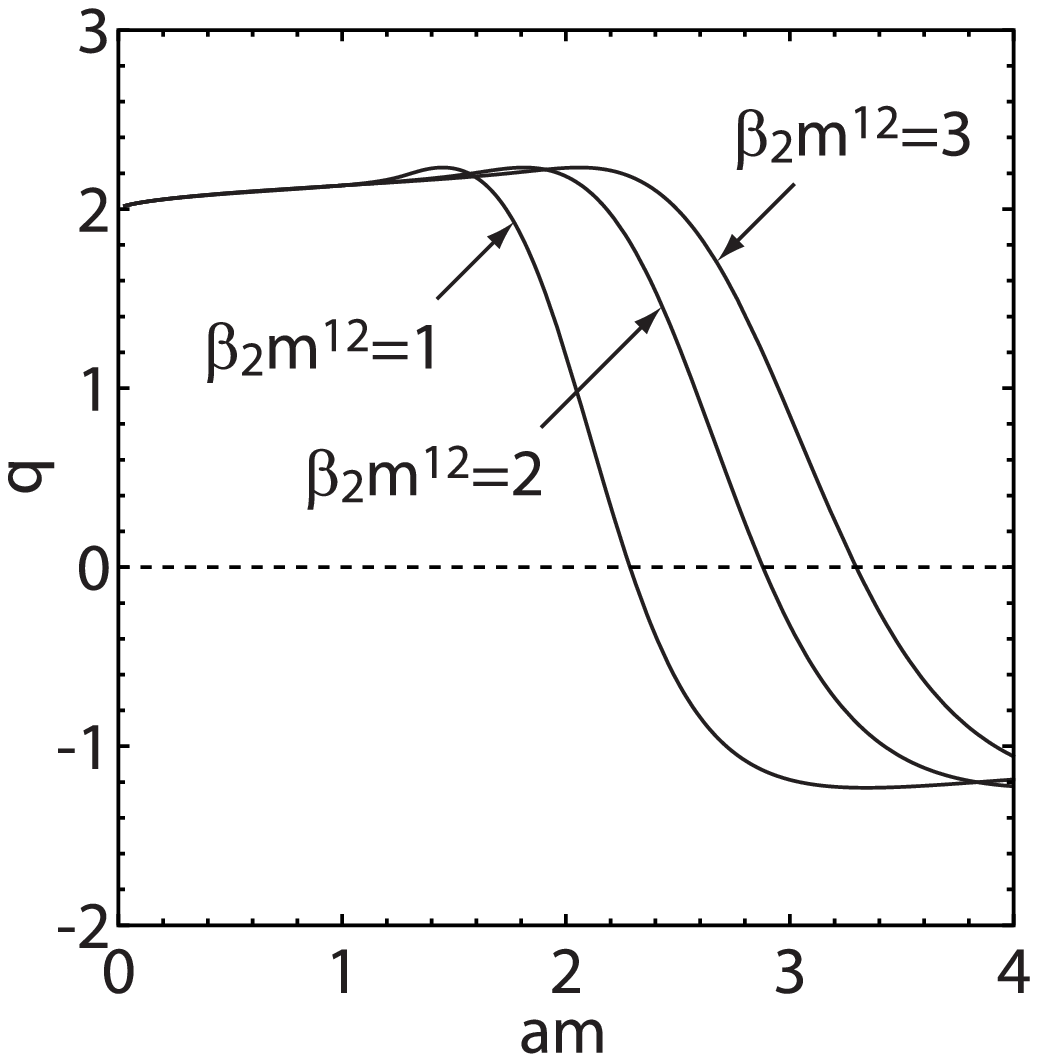}
}\\
(b) $q$
\end{center}
\end{minipage}
\caption{The behavior of the Hubble and the deceleration 
parameters for 
$\Lambda/(8\pi G m^4)=0.1$,
$\langle\bar{\psi}\psi\rangle (t=0)=m^3, a(t=0)=m^{-1}$, 
$\alpha_1 m^2=10$, $n_{max}=1$, $\beta_0=1$, 
$\beta_1m^6=-1$, $\beta_2m^{12}=1, 2, 3$ and $\beta_m=0$ 
for $m\neq 0,1,2$.}
\label{Fig:hubble02}
\end{figure}

We again assume that the non-trivial solution is realized for some of spinor 
fields and numerically solve the Friedman equation (\ref{ein:00:2}) 
under the non-trivial solution (\ref{sol:barpsipsi2}). 
The results depend
on the parameters $\alpha_n$ and $\beta_n$. Varieties of evolution can be
reproduced by choosing these parameters. Here we set finite values for
$\alpha_1$ and $\beta_n$ with $n\in\{-1,0,1,2\}$. To avoid the upper limit
for the scale factor we introduce a cosmological constant.

As is seen in Figs.~\ref{Fig:hubble01} and \ref{Fig:hubble02} 
the acceleration
period, $q<0$, appears later 
for a larger $\beta_1$ and $\beta_2$. 
If we introduce a negative
$\alpha_2$ with a positive $\alpha_3$, a similar
behavior with Fig.~\ref{Fig:hubble4} is observed
for a small $a$.
Therefore an early-time
acceleration can be
induced by the negative $\alpha_2$. Non-canonical
kinetic terms can control when the
late-time acceleration period
starts with the model parameters tuning properly.

\begin{figure}[htb]
\begin{minipage}{0.49\hsize}
\begin{center}
\resizebox{0.95\textwidth}{!}{
\includegraphics{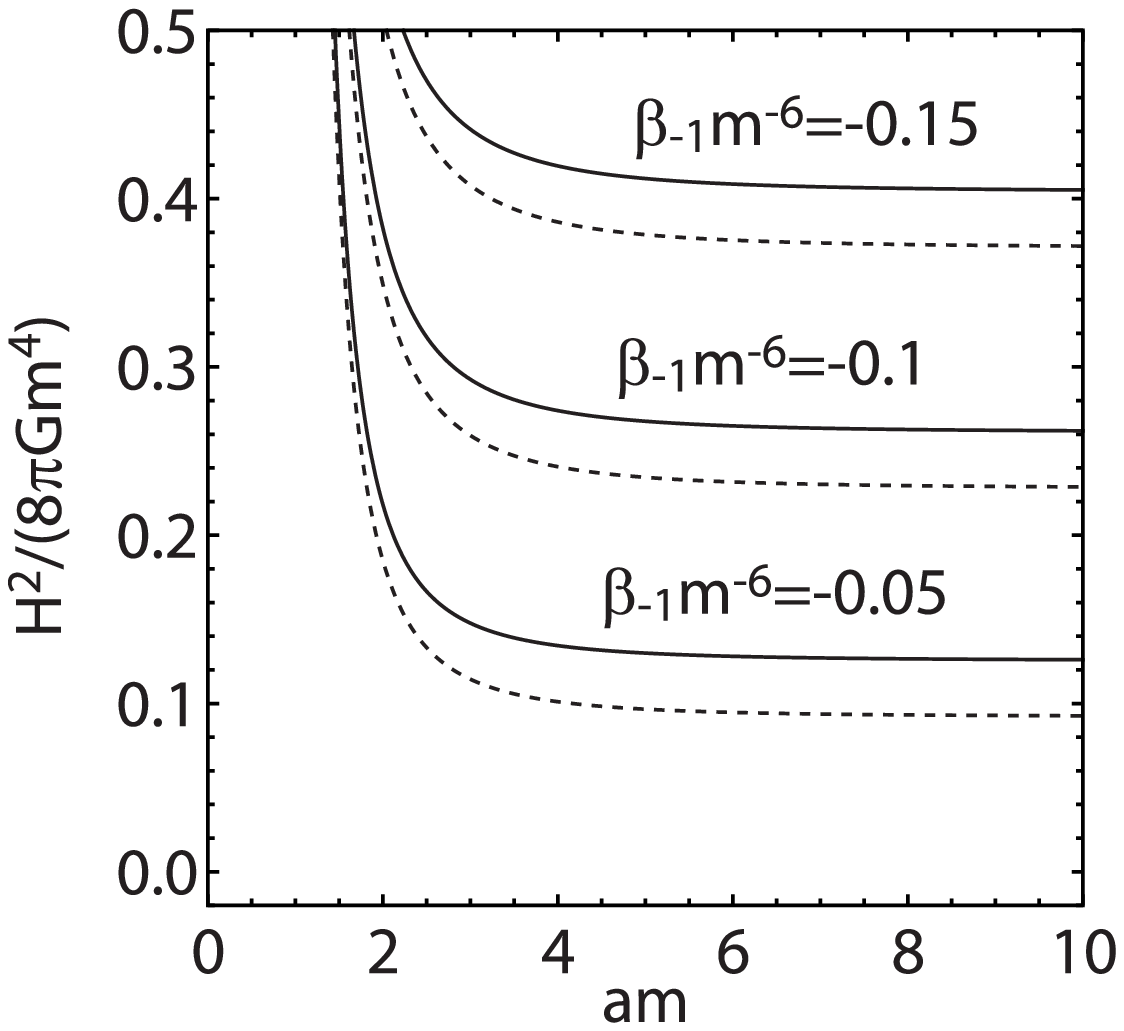}
}\\
(a) $H^2/(8\pi G m^4)$
\end{center}
\end{minipage}
\begin{minipage}{0.49\hsize}
\begin{center}
\resizebox{0.86\textwidth}{!}{
\includegraphics{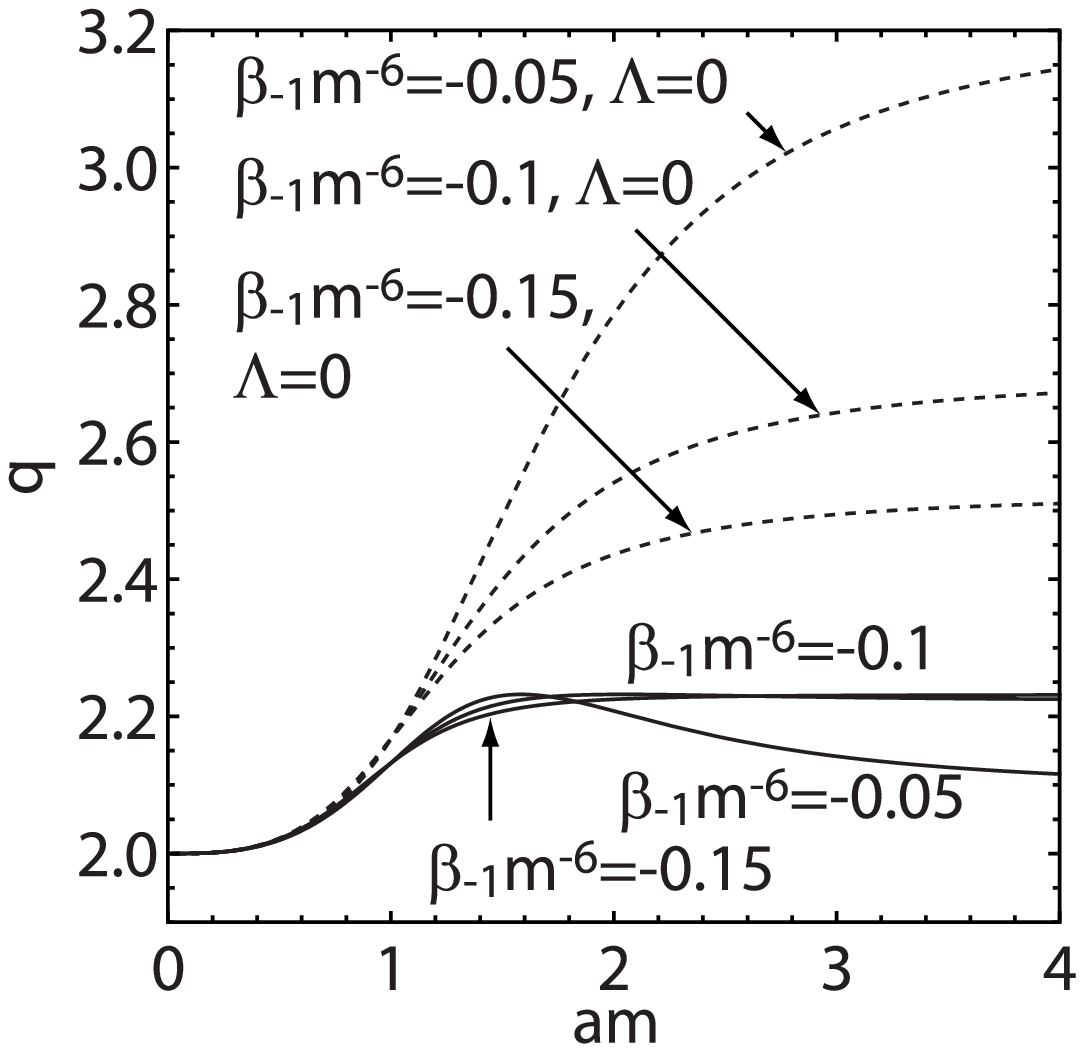}
}\\
(b) $q$
\end{center}
\end{minipage}
\caption{The solid lines show the
behavior of the Hubble and the deceleration 
parameters for $\Lambda/(8\pi G m^4)=0.1$, $\langle\bar{\psi}\psi\rangle (t=0)=m^3, a(t=0)=m^{-1}$, 
$\alpha_1 m^2=10$, $n_{max}=1$, $\beta_0=1$, 
$\beta_{-1}m^{-6}=-0.05, -0.1, -0.15$ and $\beta_m=0$ 
for $m\neq -1,0$.
The dashed lines represent the behavior for 
$\Lambda =0$.}
\label{Fig:hubble03}
\end{figure}

The term, $\beta_{-1} (\bar{\psi}\psi)^{-2}$, raises up 
the Hubble parameter. As is shown in Fig.~\ref{Fig:hubble03},
an open universe can be realized without any cosmological
constant like gravitational source for a negative $\beta_{-1}$.

\section{Condensation of spinor fields}
In QCD like theory the gauge interaction 
becomes stronger
at low energy. The strong gauge interaction induces the 
condensation of spinor fields below a critical scale, 
$a_{cr}$ or a critical temperature $T_{cr}$, see
for example Ref.~\cite{Inagaki:1997kz}. 
After the condensation the expectation value for 
the composite operator constructed by fermion and anti-fermion, 
$\bar{\psi}\psi$, develops a non-vanishing value. It is 
known as the spontaneously chiral symmetry breaking result in 
QCD \cite{Nambu:1961}. 
Thus it is natural to use the trivial solution
$\langle \bar{\psi}\psi \rangle=0$ before the universe
arrives the critical scale, $a(t)<a_{cr}$.

In a static background metric the multi-fermion
interaction is evaluated in 
Refs.~\cite{Hayashi:2008bm,Inagaki:2010py,Hayashi:2010ru}.
In the FRW universe the critical scale have 
to be fixed by observing the effective action of 
the theoretical model. For this purpose the 
quantum field theory in an unstable background metric is
necessary. It seems to have many 
difficulties \cite{Inagaki:2005qp,Inagaki:2006ff}. 
We take the remaining  problem to 
fix the critical scale as future works. However, 
the symmetry breaking changes the solution for the 
equation of motion. It has played a decisive role 
for the evolution of the universe. Here we set
the critical scale by hand and discuss the contribution
from the symmetry breaking effects at certain scales.

We suppose that the first order phase transition takes
place, the solution for the equation of motion suddenly 
jumps from the trivial to the non-trivial one.
Under the trivial solution the Hubble parameter is
constant, which corresponds to the well known de Sitter 
universe phase. A static universe realized without any 
cosmological constant like terms. The cosmological 
constant can exponentially expand the universe.
After the universe arrives at the critical scale, the
expectation value for the composite operator 
$\bar{\psi}\psi$ decelerates the cosmic expansion. 

\begin{figure}[htb]
\begin{minipage}{0.49\hsize}
\begin{center}
\resizebox{0.95\textwidth}{!}{
\includegraphics{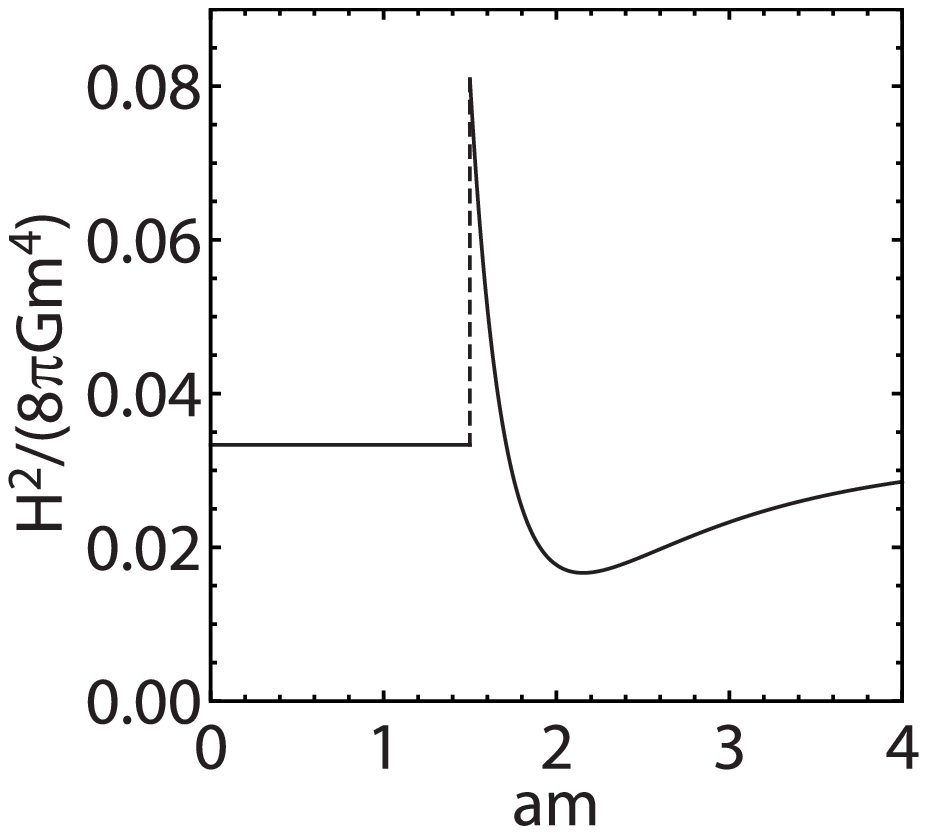}
}\\
(a) $H^2/(8\pi G m^4)$
\end{center}
\end{minipage}
\begin{minipage}{0.49\hsize}
\begin{center}
\resizebox{0.86\textwidth}{!}{
\includegraphics{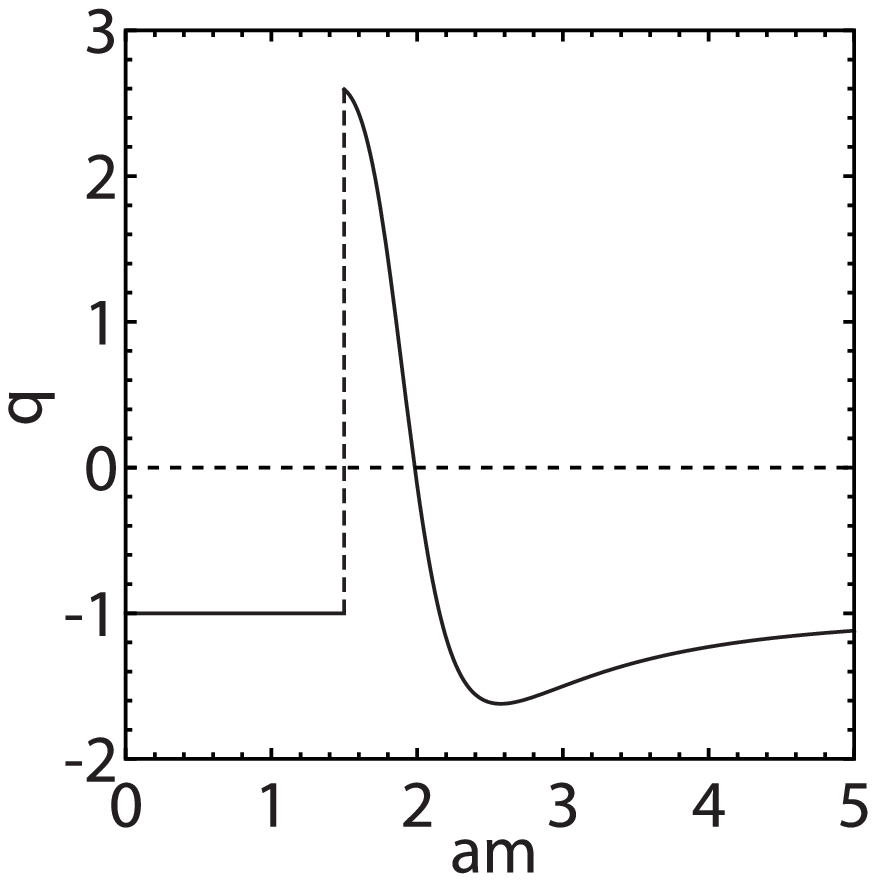}
}\\
(b) $q$
\end{center}
\end{minipage}
\caption{Behavior of the Hubble and the deceleration 
parameters for $a_{cr}m=1.5$, $\Lambda/(8\pi G m^4)=0.1$, $C=-1$, $\alpha_1 m^2=5$ and $n_{max}=1$.}
\label{Fig:hubble5}
\end{figure}

We set the critical scale at $a_{cr}m=1.5$ by hand
and draw a typical behavior of the Hubble and
the deceleration parameters for the same parameter
setting with a line for $\alpha_1 m^2=5$ in 
Fig.~\ref{Fig:hubble2}. As is seen in Fig.~\ref{Fig:hubble5}, 
the Hubble and the deceleration parameters are constant, 
$H^2=\Lambda/3$ and $q=-1$ below the critical scale $a<a_{cr}$. 
We observe the same solution with Fig.~\ref{Fig:hubble3}
above the critical scale. In this case the spinor field 
contributes in the intermediate scale. The cosmological 
constant dominated universe can be realized at the early 
and the late stage of the universe evolution.
We note that the cosmological parameters moves smoothly 
at the critical scale if the phase transition is of 
the second order.

\section{Conclusions}
We have investigated the self-interacting spinor fields as
a gravitational source
for space-time evolution via the Einstein's gravity.
It has been assumed that the 
background metric is homogeneous and isotropic in the large scale. 
We have
solved the Einstein-Dirac equations for the spinor bilinear,
$\bar{\psi}\psi$, in the FRW metric. There are two solutions.
One of the solutions, $\bar{\psi}\psi=0$, has played no significant
role for the space-time evolution.

First, we have evaluated the non-trivial solution and 
showed some
behaviors of the cosmological parameters. It has been 
observed that a self-interacting potential can induce
the early time cosmic acceleration. It is consistent with 
result in Ref.~\cite{Armendariz-Picon:2003}.
The expectation value for the mass term dominates the vacuum 
energy at a late stage of the universe evolution. Because of a negative 
expectation value for the spinor bilinear operator, $\bar{\psi}\psi$, 
mass term decelerates cosmic expansion. Only a closed universe 
can be realized without another gravitational source. 
In this work we have
introduced a cosmological constant term. The cosmological 
constant can promote the accelerated expansion of the Universe 
at a late stage. It is shown that the non-canonical kinetic term
controls when the late time acceleration starts.

Non-vanishing expectation value for the spinor bilinear breaks
the chiral symmetry for the spinor field. 
It is believed that
high temperature, high
density and strong curvature may restore the broken symmetry at 
the early universe. Thus it is not always valid to adopt the 
non-trivial solution of the Einstein-Dirac equations. 
The trivial solution is realized before the critical scale,
$a<a_{cr}$. Here we have set a critical scale, $a_{cr}$, by hand 
and demonstrate the possibility that the spinor fields can 
decelerate the cosmic expansion in an intermediate scale.

It should be noted that the late time accelerated expansion 
or the dark energy phenomena can be produced without a 
cosmological constant. An interesting mechanisms 
is found in a modified gravity. For 
general review of the later time accelerated
expansion in the modified gravity, see Ref.~\cite{Nojiri:2010wj}. 
It is not hard to take account of spinor source in above scheme.

In this paper we restrict the potential as a function of
only the scalar invariant operator, $\bar{\psi}\psi$, and consider 
isotropic universe evolution for large scale. 
Other types of spinor bilinear may 
generate an angular momentum. It is known that an anisotropic 
universe is generally realized due to the spinor angular momentum. 
Hence, it is interesting to include other types of spinor bilinear 
and study the model in an anisotropic universe.
We have not calculated the critical scale, $a_{cr}$, 
starting from the action. 
It should be fixed by observing the stationary point 
of the effective action. To evaluate the effective action we 
should take into account the quantum and thermal effects in an
expanding universe background. We will continue our work further
and hope to report on these problems.

\section*{Aknowledgment}
This work is partly supported by Nature Science Fondation of China (NSFC) under 
contract No.11075078 and 10675062.

\end{document}